# Dipolar Magnetic Interactions and A-type Antiferromagnetic Order in the Zintl Phase Insulator EuZn$_2$P$_2$

Tanya Berry,[1,2*] Veronica J. Stewart,[1,2] Benjamin W. Y. Redemann,[1,2] Chris Lygouras,[2] Nicodemos Varnava,[3] David Vanderbilt,[3] Tyrel M. McQueen[1,2,4**]

[1]Department of Chemistry, Johns Hopkins University, Baltimore, Maryland 21218, USA

[2]Institute for Quantum Matter, William H. Miller III Department of Physics and Astronomy, Johns Hopkins University, Baltimore, Maryland 21218, USA

[3]Department of Physics & Astronomy, Rutgers University, Piscataway, New Jersey 08854, USA

[4]Department of Materials Science and Engineering, Johns Hopkins University, Baltimore, Maryland 21218, USA

* Email: tberry@ucdavis.edu and **Email: mcqueen@jhu.edu

**Abstract:** Zintl phases, containing strongly covalently bonded frameworks with separate ionically bonded ions, have emerged as a critical materials family in which to couple magnetism and strong spin orbit coupling to drive diverse topological phases of matter. Here we report the single-crystal synthesis, magnetic, thermodynamic, transport, and theoretical properties of the Zintl compound EuZn$_2$P$_2$ that crystallizes in the anti-La$_2$O$_3$ $P$-3m1 structure, containing triangular layers of Eu$^{2+}$ ions. In-plane resistivity measurements reveal insulating behavior with an estimated band gap of $E_g$=0.11eV. Specific heat and magnetization measurements indicate antiferromagnetic ordering at $T_N$=23 K. Curie-Weiss analysis of in-plane and out-of-plane magnetic susceptibility from $T$=150-300K yields $p_{eff}$=8.61 for $\mu_0H\perp$c and $p_{eff}$=7.74 for $\mu_0H//$c close to the expected values for the $4f^7$ $J$=$S$=7/2 Eu$^{2+}$ ion and indicative of weak anisotropy. Below $T_N$, a significant anisotropy of $\chi_\perp/\chi_{//}\approx$2.3 develops, consistent with A-type magnetic order as observed in isostructural analogs and as predicted by the density functional theory calculations reported herein. The positive Weiss temperatures of $\theta_W$=19.2 K for $\mu_0H\perp$c and $\theta_W$=41.9 K for $\mu_0H//$c show a similar anisotropy and suggest competing ferromagnetic and antiferromagnetic interactions. Comparing Eu magnetic ordering temperatures across trigonal Eu$M_2X_2$ (M=divalent metal, X=pnictide) shows that EuZn$_2$P$_2$ exhibits the highest ordering temperature, with variations in $T_N$ correlating with changes in expected dipolar interaction strengths within and between layers and independent of the magnitude of electrical conductivity. These results provide experimental validation of the crystochemical intuition that the cation Eu$^{2+}$ layers and the anionic ($M_2X_2$)$^{2-}$ framework can be treated as electronically distinct subunits, enabling further predictive materials design.

**Introduction:** Zintl phase materials are known to exhibit various phenomena including topological insulators, Dirac semimetals, colossal magnetoresistance, anomalous Hall effect, and spin Hall effect, with applications ranging from thermoelectricity, superconductivity, catalysis, spintronics, optoelectronics, and solar cells.[1-15] Zintl phases typically consist of a strongly covalently bonded anionic framework with separate cations to provide charge balance. Due to this difference in bonding types, the cationic and anionic subunits can often be tuned or modified independently, enabling systematic materials design of phenomena that requires precise tuning of multiple factors simultaneously.[14] A recent example of the success of this tunability is the discovery of magnetic topological insulators in EuIn$_2$As$_2$ and MnBi$_2$Te$_4$, where the electronic structure of the anionic framework provides the topology, while the magnetic order comes from distinct electrons on the cations.[1,6,16]

When the cationic framework is constructed of magnetic ions and the anionic framework is non-magnetic with a closed-shell electron configuration, Zintl phases can realize a variety of magnetic lattices and enable concrete comparisons to analytical theories.[17] Further, adjusting the electronic structure of the anionic framework with a fixed cationic lattice enables determination of the key interaction pathways within and between structural subunits, critical knowledge needed to discover and engineer novel electronic and magnetic phases.

More concretely, layered, insulating, Eu-based Eu$M_2X_2$ (M=metal, X=pnictide) Zintl compounds provide realizations of weakly coupled square or triangular magnetic lattices, depending on the preferred geometry of the $M_2X_2$ layer (ThCr$_2$Si$_2$-type or anti-La$_2$O$_3$-type respectively). The half-filled, $L=0$ nature of Eu$^{2+}$ ($J=S=7/2$) ensures minimal effects from local crystal fields (single ion anisotropy). In such materials, Eu can be partially or fully substituted with Ba, Sr, and/or Ca to dilute the magnetic lattice while holding the overall unit cell parameters constant. Alternatively, the anionic framework can be adjusted to change the cation ion separation and hence strength of the magnetic interactions within or between cationic layers. By adjusting the electronic structure of the anionic framework, the degree of interaction screening (or enhancement due to, e.g., superexchange or RKKY interactions) is also variable.

In this work, we report the physical properties of EuZn$_2$P$_2$ single crystals prepared via flux growth, and provide comparisons to other members of the family. Single-crystal x-ray diffraction at $T=213$ K confirms the previously reported anti-La$_2$O$_3$ structure type, with triangular Eu$^{2+}$ layers separated by (Zn$_2$P$_2$)$^{2-}$ layers of edge-sharing tetrahedra.[18] Bulk resistivity measurements reveal insulating behavior with an estimated band gap of $E_g$=0.11eV. Specific heat measurements reveal a lambda anomaly at $T_N = 23$ K, with a broad hump continuing down to $T=2$ K. After fitting a phonon model to $T<25$ K, the estimated integrated change in recovered magnetic entropy is $\Delta S_{mag}$=11.3 Jmol$^{-1}$K$^{-1}$ from $T=0$ K to $T=T_N$, close to ⅔ of the total entropy recovered for $S=7/2$ ions, $\Delta S_{mag}=R\ln(8)$. Magnetization measurements show a cusp at $T_N$=23 K, indicative of the formation of Eu antiferromagnetic order. Above $T_N$, Curie-Weiss analysis yields $p_{eff}$=8.61 for $\mu_0H\perp c$ and $p_{eff}$=7.74 for $\mu_0H//c$, close to the expected value for 4$f^7$ $J=S=7/2$ Eu$^{2+}$ ions ($p_{eff}$=7.94). Below $T_N$, a significant anisotropy of $\chi_\perp/\chi_{//}\approx 2.3$ develops, consistent with A-type magnetic order as observed in isostructural analogs. The positive Weiss temperatures of $\theta_W$=19.2 K for $\mu_0H\perp c$ and $\theta_W$=41.9 K for $\mu_0H//c$ show a similar anisotropy, and suggest competing ferromagnetic and antiferromagnetic interactions. Density functional theory (DFT) calculations recapitulate the insulating behavior and predict A-type Eu magnetic order, in agreement with our observations. Overall, such results allow us to expand our knowledge in the chemical perspectives in the nature of magnetism in Eu-based Zintl compounds and its correlation to the electrical conductivity.

**Results and Discussion:**

**Structure:** The single-crystal x-ray diffraction (SCXRD) data for EuZn$_2$P$_2$ at $T=213$ K are consistent with the previously reported structure with the trigonal space group, $P$-3m1.[19] A precession image along the (hk0) plane is shown in Fig. 1a. Tables 1, 2, and 3 provide the details, atomic coordinates and atomic displacement parameters of the final refinement, which converged to R1=1.04% and wR2=2.46%. The bond valence sum for Eu is 1.6+ implying that the Eu is divalent, i.e. Eu$^{2+}$.[19] As-grown crystals of EuZn$_2$P$_2$, Figure. 1b, were found to be phase pure via powder x-ray diffraction after cleaning the excess tin from surfaces. The Zintl structure of EuZn$_2$P$_2$, Figure. 1c, contains Eu$^{2+}$ layers separated by an anionic framework of [Zn$_2$P$_2$]$^{2-}$ edge-sharing tetrahedra. The stacking of these layers forms an octahedral coordination environment for Eu$^{2+}$.

The structure of EuZn$_2$P$_2$ is a member of the La$_2$O$_3$ family, sometimes called the Ce$_2$O$_2$S or CaAl$_2$Si$_2$ type, and structurally written as [La$_2$O$_2$]$^{2+}$O$^{2-}$.[18,20] EuZn$_2$P$_2$ is formally of the anti-La$_2$O$_3$-type, with anionic [Zn$_2$P$_2$]$^{2-}$ layers and cationic Eu$^{2+}$. Looking beyond those containing Eu, this is a very common structure type, including binaries like Mg$_3$Sb$_2$ (Figure. 1d) ternaries such as Li$_2$MnO$_2$ (Figure. 1e) and quaternaries such as NaLiCdS$_2$ (Figure. 1f).[20-24] All these structures crystallize into the same $P$-3m1 space group rationalized by the Zintl concept where the cation donate their valence electrons to stabilize the anionic tetrahedral framework. In these structures, individual tetrahedra are polar – with an elongation along the stacking axis (c-axis). For example, in the case of EuZn$_2$P$_2$, one Zn-P bond being 2.5303(10) Å, but the three equivalent Zn-P bonds being 2.4608(3) Å. These units are arranged in an alternating up/down pattern so there is no net electric polarity even in a single layer. However, the anionic layers might be expected to have a significant electric quadrupolar response, something that appears to be uninvestigated to date and is worthy of future study.

**Heat Capacity:** As Eu magnetic order is expected, heat capacity measurements were carried out on EuZn$_2$P$_2$ single crystals from $T$=2-300 K to identify possible phase transitions. A lambda anomaly at $T_N$ =23 K is observed, Figure. 2a, and, when combined with magnetization (vida infra), is indicative of Eu antiferromagnetic order. The overall shape of the transition – a sharp peak followed by a broad hump at lower temperatures, is commonly observed in $S$=7/2 magnets on simple-cubic-like lattices.[24] Extraction of the magnetic contribution to the specific heat is non-trivial, as the most direct non-magnetic structural analog would be BaZn$_2$P$_2$, which adopts the distinct ThCr$_2$Si$_2$ structure type, whereas others have large differences in molar masses relative to EuZn$_2$P$_2$.[25] We thus chose to construct a phonon model to fit the high temperature specific heat. The chosen model, the two Debye model is given by:

$$\frac{C_p}{T} = \frac{C_D(\theta_{D1}, s_1, T)}{T} + \frac{C_D(\theta_{D2}, s_2, T)}{T} \quad (1)$$

$$C_D(\theta_D, T) = 9sR \left(\frac{T}{\theta_D}\right)^3 \int_0^{\theta_D/T} \frac{(\theta/T)^4 e^{\theta/T}}{[e^{\theta/T}-1]^2} d\frac{\theta}{T} \quad (2)$$

Where $\theta_{D1}$ and $\theta_{D2}$ are the Debye temperatures, $s_1$ and $s_2$ are the oscillator strengths, and R is the molar Boltzmann constant. It is physically motivated by the distinct subunits: the phonon modes of the anion framework are expected to be distinct, energetically, from those arising from the cationic lattice. The model parameters from the least-squares refinement to the data for $T$ >25 K, are given in Table 4. The total oscillator strength $s_1+s_2$ = 5.18(6). This is in good agreement with the expected value of 1+2+2 = 5, the total number of atoms per formula unit in EuZn$_2$P$_2$.

After subtracting this phonon contribution, the sample heat capacity from $T$=2-300K was integrated to determine the change in entropy corresponding to magnetic order in EuZn$_2$P$_2$. Figure. 2b shows that the change in the magnetic entropy reaches a maximum of $\Delta S_{mag}$=11.3 J.mol$^{-1}$.K$^{-1}$ just above $T_N$. A compound that consists of Eu$^{2+}$ without any multivalence of Eu or other magnetic ions should have a recovery of entropy $\Delta S_{mag}$=$R$ln(8)=17.3 J.mol$^{-1}$.K$^{-1}$, so the observed value is ~⅔ of that expected. The most likely explanation is that the two Debye model is over-estimating the phonon contribution, particularly in the region just above $T_N$, as a gradual continued entropy recovery is expected for $S$=7/2 beyond $T_N$.[24] The unphysical dip of the integrated $\Delta S_{mag}$ above $T_N$ in Figure. 2b is consistent with this possibility. However, another possibility that cannot be ruled out is that there is additional entropy below *T*= 2K that is not captured by a linear extrapolation of *Cp/T* approaching 0 at *T*=0 K.

**Magnetization:** Magnetization measurements provide details on the type of magnetic order in EuZn$_2$P$_2$ and enable determination of the magnetic phase diagram. Temperature-dependent magnetic susceptibility, measured with an applied field perpendicular and parallel to the layers (c axis), Figures. 3a and 3b, show a clear antiferromagnetic phase transition at $T_N$=23K. The data from *T*=75-300K for $\mu_0H\perp$c and *T*=150-300 K for $\mu_0H$//c are well described by the Curie-Weiss law:

$$\chi = \frac{C}{T-\theta_{cw}} + \chi_0 \quad (3)$$

Where $\chi_0$ is the temperature independent susceptibility, C is the Curie constant, and $\theta_W$ is the Weiss temperature. The fits are shown in Figure 3d and 3e and the best least-squares parameters are given in Table 5. The $p_{eff}$ extracted from the Curie constants are 8.61 for $\mu_0H\perp$c and 7.75 for $\mu_0H$//c, are consistent with the theoretical $p_{eff}$ =7.94 for Eu$^{2+}$. The positive Weiss constants indicate dominant ferromagnetic interactions; in combination with the observed antiferromagnetic order, this implies a mixture of magnetic exchange interactions. Another noticeable feature, in the susceptibility data is the anisotropy in $\mu_0H\perp$c and $\mu_0H$//c, Figure 3c and 3f. The anisotropy is larger below the $T_N$ in comparison to above the $T_N$ with $\chi_\perp/\chi_{//}$ in the range 0.8-1.25 above T$_N$, rising rapidly below $T_N$ to $\chi_\perp/\chi_{//}$~2.3 at *T*=2K. This anisotropy is not due to demagnetization – which has been corrected for in these data – and instead implies spin alignment in the ab plane below $T_N$. This is in agreement with observations on related Eu compounds.[26-29]

To further evaluate the magnetic behavior, M(H) curves were collected as a function of temperature for both $\mu_0H\perp$c and $\mu_0H$//c, Fig. 4a and 4b. In both crystal directions, the magnetization saturates at 7$\mu_B$, the expected value for divalent Eu$^{2+}$. This saturation rules out any Eu$^{2+}$/Eu$^{3+}$ mixed valency in EuZn$_2$P$_2$. The highly linear M(H) behavior before saturation in both directions is common to many Eu$^{2+}$ materials.[26-29] However, there are noticeable differences in the M(H) data in the $\mu_0H\perp$c and $\mu_0H$//c directions. The saturation of 7$\mu_B$ happens at different magnetic fields, which implies that there is a difference in the magnetic stiffness in the two directions. These anisotropic features are consistent with the susceptibility results in Figure. 3a-f. It takes little applied magnetic field ($\mu_0H$=0.75 T) to fully polarize the spins in the $\mu_0H$//c direction in comparison to the $\mu_0H\perp$c direction ($\mu_0H$=2.25 T). Another feature is that the Eu spins in the EuZn$_2$P$_2$ do not undergo any apparent phase transitions as a function of field and smoothly transition into the field polarized state. To extract the characteristic fields associated with these behaviors, Figures. 4c and 4d show the derivative as a function of the field.

Figure. 5a and 5b show the critical fields in M(H) as a function of field direction. From the M(T) measurements in Figure 3a and 3b, we know that EuZn$_2$P$_2$ is net antiferromagnetic. The Eu spins then reorient themselves to a field polarized paramagnetic/ferromagnetic state on applying a sufficiently large magnetic field. These data are sufficient to identify the type of magnetic order present: as described above, the spins lie in the ab plane. The order cannot be G-type (all antiferromagnetic within and between Eu layers), since in that case a proper phase transition should be observed with an applied field. The triangular lattice within each layer makes C-type order (ferromagnetic between layers, antiferromagnetic within layers) unlikely as there is no evidence of a structural phase transition. Further, with C-type order one would expect a field-driven magnetic transition along one of the two directions, while none are observed. We are thus left with A-type order (ferromagnetic layers stacked antiferromagnetically) as the prime candidate for the magnetic order. Such an order can continuously rotate to a field polarized state independent of the applied field direction, in agreement with the data here. It is also in agreement with the known magnetic order in isostructural analogs.[30-33]

**Resistivity:** To explore the electronic behavior, the temperature dependence of the electrical resistivity was measured for $T$=140-400K, Figure 6a. The resistivity decreases with temperature in the paramagnetic state and the order of magnitude of the resistivity is $10^3$ $\Omega$.cm at $T$=300 K consistent with insulating behavior in $EuZn_2P_2$. In order to extrapolate the bandgap through the resistance, the Arrhenius model was fitted to the resistivity data as $ln(\rho/\rho_{T=400K})$ versus $T^{-1}$ in Figure 6b. In the Arrhenius model, a linear trend explains that the resistivity follows an activation type relation, $\rho \propto e^{(Eg/kBT)}$, and fits $EuZn_2P_2$ well over the range measured. The extracted bandgap is $E_g$=0.11 eV. We note that a recent conference proceedings claims to have observed similar behavior in $EuZn_2P_2$.[34]

**Theory:** DFT calculations were performed for all four simple magnetic configurations with spins in the ab plane: G-type, C-type, A-type, and all ferromagnetic (FM). The A-type order, Figure 7a, was found to be lowest in energy. This is in agreement with the magnetization measurements and provides further evidence of A-type order in $EuZn_2P_2$. DFT also predicts $EuZn_2P_2$ to be an electrical insulator in the antiferromagnetic state. This insulating behavior persists in the FM configuration, Figure 7b, and implies that $EuZn_2P_2$ should be an insulator at all temperatures, in agreement with experiment (we note that the gap for comparison to theory is 2Eg as defined here, i.e. 0.22 eV). These calculations also demonstrate how the distinct subunits of a Zintl phase couple: the direct bandgap at $\Gamma$ is $E_g$=0.60 eV in the antiferromagnetic configuration and $E_g$=0.48 eV in the ferromagnetic configuration (a 20% change), and the indirect bandgap from $\Gamma$ to M is $E_g$=0.48 eV in the antiferromagnetic configuration and $E_g$=0.30 eV in the ferromagnetic configuration (a 38% change). This large change in band gap occurs even though the states at the top of the valence band and the bottom of the conduction band are derived from the non-magnetic Zn and P atoms (and not Eu) – Eu 4$f$ states are 1.2 eV below the valence band maximum in both cases. Thus the large changes in predicted gap thus reflects a large exchange splitting of the Zn-P bands due to the Eu magnetism.

Given the impact of the Eu magnetism on the electronic structure of the $Zn_2P_2$ framework, it is natural to ask if the reverse is true – does the electronic structure of the anion framework significantly impact the magnetic order of the Eu lattice? To address this question, Table 6 gives key parameters for a set of isostructural Eu$M_2X_2$ (M = metal, X = pnictide) magnets, including intra- and inter- layer distances, and room temperature resistivities. Note that in all cases, an insulating band structure (with either a positive or negative – i.e. semimetal – gap) is expected on the basis of electron count and bonding. Several trends are immediately apparent. First, $EuZn_2P_2$ has a higher Eu ordering temperature than any other known member of this family. Second, for a given metal M, the materials become more conductive and have a lower Eu ordering temperature as X increases down the pnictogen column. The magnitudes of the changes, however, are not uniform for different M. Given the large $J=S$=7/2 for Eu$^{2+}$, if the Eu-Eu interactions are primarily dipolar in nature, then one expects the magnetic interaction strength, and hence $T_N$, to scale as $1/d^3$, where d is the interion separation. We note that this holds even if the $M_2X_2$ layers screen a portion of the dipolar interaction, as would be expected in general when the magnetic permeability is not exactly unity.[42] Simply using the intralayer or interlayer distances gives partial trends, but does not unify the observations across different choices of M. We thus applied a multilinear regression to determine if some linear combination of $1/d_{nn}^3$ ($d_{nn}$ = intralayer Eu-Eu distance) and $1/d_{il}^3$ ($d_{il}$ = interlayer Eu-Eu distance) could explain all the data in Table 6. This yielded coefficients close to a 1:-4 ratio. Fixing the combination at exactly this ratio yields the trend in Figure 8, which shows a monotonic dependence of $T_N$ on this mixture of Eu-Eu distances across different choices of M. This relationship holds across five orders of magnitude of conductivity, and demonstrates that the Eu-Eu interactions are dipolar in nature and that the primary effect of the anionic framework is to modulate the dipolar interaction via changes in Eu-Eu distances (rather than, e.g., being due to superexchange or RKKY interactions).[43]

**Conclusion:** $EuZn_2P_2$ single crystals prepared via a flux technique crystallize in the trigonal anti-$La_2O_3$ structure type and exhibit Eu antiferromagnetic order at $T_N$= 23 K. Heat capacity results showed a change in the magnetic entropy saturation at ~⅔Rln(8) likely limited by the available phonon contribution model. Via magnetic susceptibility measurements the anisotropy of the $\mu_0H\perp c$ and $\mu_0H//c$ directions of $EuZn_2P_2$ single crystals was characterized. A significant anisotropy develops below $T_N$. The magnetic phase diagram allows identification of the magnetic structure of $EuZn_2P_2$ as A-type magnetic order. Resistivity measurements indicate insulating behavior with an experimental bandgap of $E_g$=0.11 eV, consistent with DFT calculations. Comparisons amongst related materials shows that the dominant effect of the anionic framework on the Eu lattice magnetism can be understood as arising from magnetic dipolar interactions. Overall, our results help us gain insights into the polyanionic contributions to the conductivity and magnetism in the $EuM_2X_2$ structures. For future work, local probes to understand the local coordination environment in $EuZn_2P_2$ would allow understanding of the structural coordination and high resistance measurements to understanding the conductivity close to Neél temperature to evaluate the conduction pathways in $EuZn_2P_2$.

**Method:** Single crystals of $EuZn_2P_2$ were synthesized using elemental Eu (ingot, Yeemeida Technology Co., LTD 99.995%), Zn (shots, Sigma-Aldrich 99.99%), red P (Sigma-Aldrich 97+%), and Sn (shots, Sigma-Aldrich 99.99%). The elements were put in a 1:2:2:45 ratio and a total mass of 5 grams were put in a Canfield crucible (size: 2mL). Zn and Sn were placed in the crucible at atmospheric conditions, while Eu and P were added later in an Ar-filled glove box. The Canfield crucible was placed in a quartz ampoule with quartz wool below and above the crucible, evacuated, and sealed under $1.2 \times 10^{-2}$ torr of pressure. The evacuated ampoules were loaded in a box furnace at an angle of 45º. The temperature was ramped at rate of 100ºC/h to $T$=500 ºC for 4 hours. This step allowed for Sn flux to be in a liquid state. The furnace was then ramped from $T$=500ºC to $T$=1150ºC at the rate of 100ºC/hr and held for 24 hr. The furnace was then slowly cooled to $T$=850ºC at the rate of 5ºC/hr, then removed hot, inverted, and immediately centrifuged. Centrifugation took 2-3 minutes. Hexagonal plate-like crystals of size 3 mm along the long direction were removed from the frit. These single crystals of $EuZn_2P_2$ were found to be stable in air over period of months.

Powder x-ray diffraction (XRD) data, used to confirm phase purity of the single crystals, were collected over an angle range of 5°–60° on a laboratory Bruker D8 Focus diffractometer with a LynxEye detector and Cu Kα radiation. Single-crystal x-ray data were collected on a Bruker-Nonius X8 Proteum (Mo Kα radiation) diffractometer equipped with an Oxford cryostream. Integration and scaling were performed using CrysalisPro (Version 1.171.39.29c, Rigaku OD, 2017). A multiscan absorption correction was applied using SADABS.[44] The structure was solved by direct methods, and successive interpretations of difference Fourier maps were followed by least-squares refinement using SHELX and WinGX.[45,46]

Magnetization data were collected on a Quantum Design Magnetic Property Measurement System (MPMS). Magnetic susceptibility was approximated as magnetization divided by the applied magnetic field ($\chi \approx M/H$). To account for shape effects, a demagnetization correction was incorporated, with $N$=0.76 for $\mu_0 H$//c and $N$=0.18 for $\mu_0 H \perp$c. In addition, heat capacity data were collected on a Quantum Design Physical Properties Measurement System using the semiadiabatic method and a 1% temperature rise.

The resistivity measurements were performed in a Quantum Design Physical Property Measurement System (PPMS) using a Lake Shore Model 372 AC Resistance Bridge. An AC current set to autorange between 10nA - 1μA with an excitation frequency of 13.7 Hz was applied. The measurement was performed on a crystal with a four-probe configuration consisting of platinum wires and Epo-Tek silver epoxy H20E-FC. The picture of the single crystal with the Pt leads can be found in Figure 6a.

Density functional theory (DFT) based first principles calculations were performed using the projector augmented-wave (PAW) method as implemented in the VASP code.[47,48] We used the PBE exchange-correlation functional as parametrized by Perdew-Burke-Ernzerhof.[49,50] For the self-consistent calculations a Monkhorst-Pack k-mesh of size 15×15×5 was used to sample the Brillouin zone (BZ). The energy cutoff is chosen 1.5 times as large as the values recommended in relevant pseudopotentials. Spin-orbit coupling (SOC) was included self-consistently. The Eu 4f states were treated by employing the GGA+U approach with the U value set to 5.0eV.

**Acknowledgments:** The authors would like to thank M. Siegler for technical assistance. This work was supported as part of the Institute for Quantum Matter and Energy Frontier Research Center, funded by the U.S. Department of Energy, Office of Science, Office of Basic Energy Sciences, under Award DE-SC0019331. The MPMS was funded by the National Science Foundation, Division of Materials Research, Major Research Instrumentation Program, under Award 1828490.

**Figure 1.** (a) Single-crystal precession image of the (hk0) plane. (b) As-grown single-crystal of EuZn$_2$P$_2$. Structures of (c) EuZn$_2$P$_2$, (d) Mg$_3$Sb$_2$, (e) Li$_2$MnO$_2$, and (f) NaLiCdS$_2$ that crystallize in *P*-3m1 (164) and are anti-La$_2$O$_3$ structures containing 2D trigonal layers of cations separated by anionic layers of edge-sharing tetrahedra. The black box represents the unit cell in each case.[20-24]

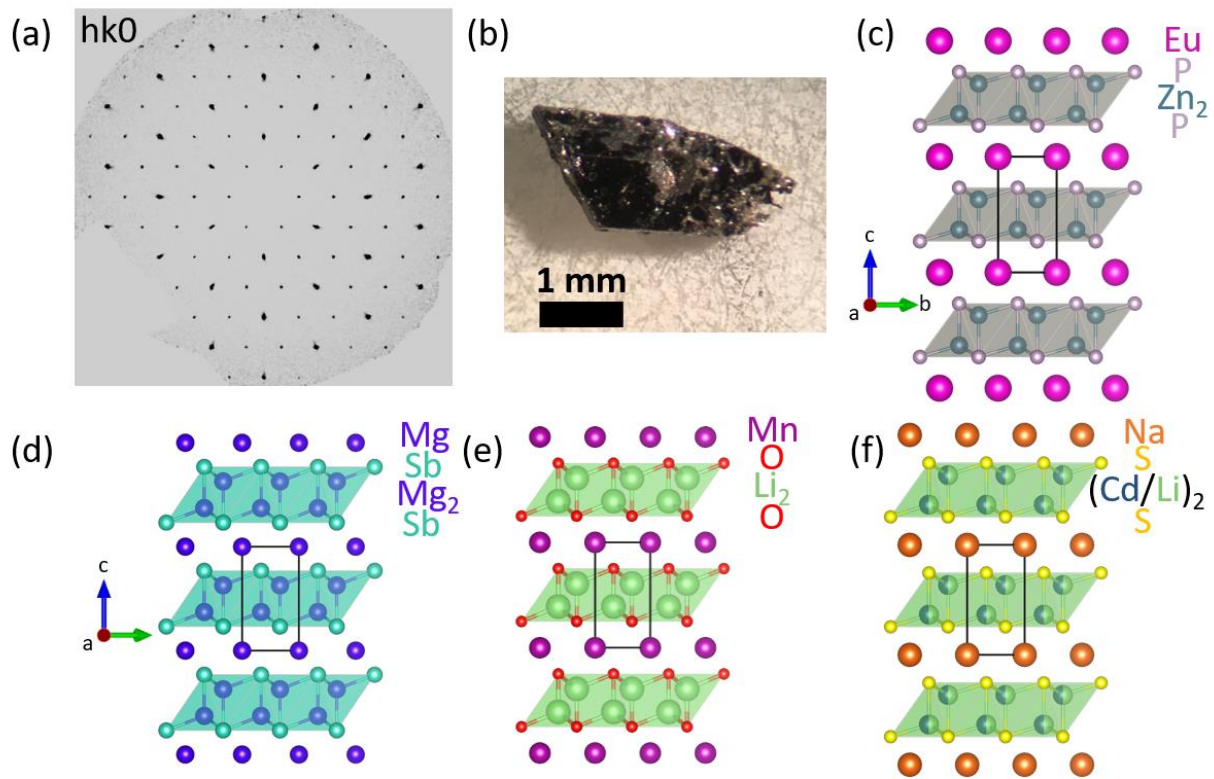

**Figure 2. (a)** Specific heat divided by temperature as a function of temperature for EuZn$_2$P$_2$ single crystals from $T$=2–300 K. The maroon filled squares denote experimental data, and the green is the phonons modeled using the Two-Debye model from $T$=30–300 K. The sharp transition at $T$=25 K is attributed to the antiferromagnetic phase transition. The bump around $T$=270 K is from Apiezon N grease used during the measurement. **(b)** The change in magnetic entropy was integrated after subtracting the phonons from $T$=2–300K. The $\Delta S$mag is close to ~2/3Rln(8). The light green in $\Delta S$mag at around $T$=270 K is coming from Apiezon N grease.

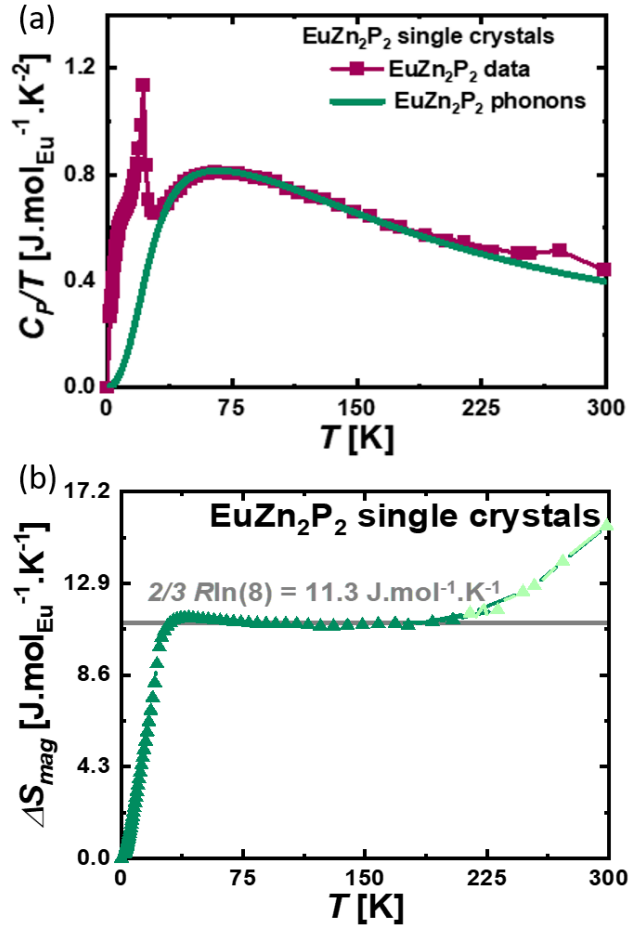

**Figure 3.** (a) Magnetization as a function of temperature with $\mu_0 H \| c$ and $\mu_0 H = 0.1$–7 T and $T=2$–300 K. The $\mu_0 H = 0.1$ T data show a clear AFM transition at $T=25$ K, with a decrease in both sharpness of the transition and the temperature of the transition as the field increases. (b) Magnetization as a function of temperature $\mu_0 H \perp c$ from $\mu_0 H = 0.1$–7 T and $T=2$–300 K. The $\mu_0 H = 0.1$ T data show a kink at $T=25$ K followed by an upturn, both of which are suppressed for $\mu_0 H > 0.1$ T. (c) Comparison in the magnetization as a function of temperature, $\mu_0 H \perp c$ and $\mu_0 H \| c$, at $\mu_0 H = 0.1$ T over $T = 2$–300 K. (d) Curie Weiss analysis for $\mu_0 H \| c$ from $\mu_0 H = 0.1$ T in the range $T=2$–300 K, (e) Curie Weiss analysis $\mu_0 H \perp c$ from $\mu_0 H = 0.1$ and $T=2$–300 K, and (f) ratio of magnetization $\mu_0 H \| c$ and $\mu_0 H \perp c$ at $\mu_0 H = 0.1$ T and $T=2$-300 K displaying the anisotropy below $T_N$.

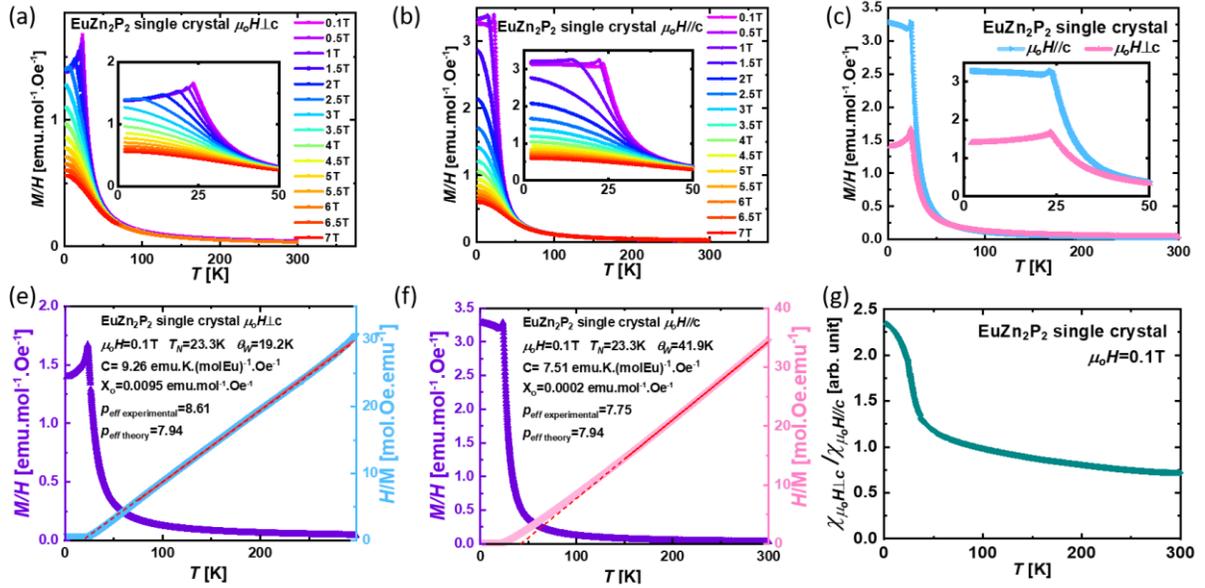

**Figure 4.** For each plot below, the following temperatures are shown: $T = 2, 5, 8, 10, 20, 21, 100$, and 300 K. **(a)** Magnetization as a function of magnetic field with $\mu_o H \| c$ from $\mu_o H = -7$ to 7 T; **(b)** Magnetization as a function of magnetic field with $\mu_o H \perp c$ from $\mu_o H = -7$ to 7; **(c)** derivative of magnetization over magnetic field as a function of magnetic field with $\mu_o H \| c$; and **(d)** derivative of magnetization over magnetic field as a function of magnetic field with $\mu_o H \perp c$.

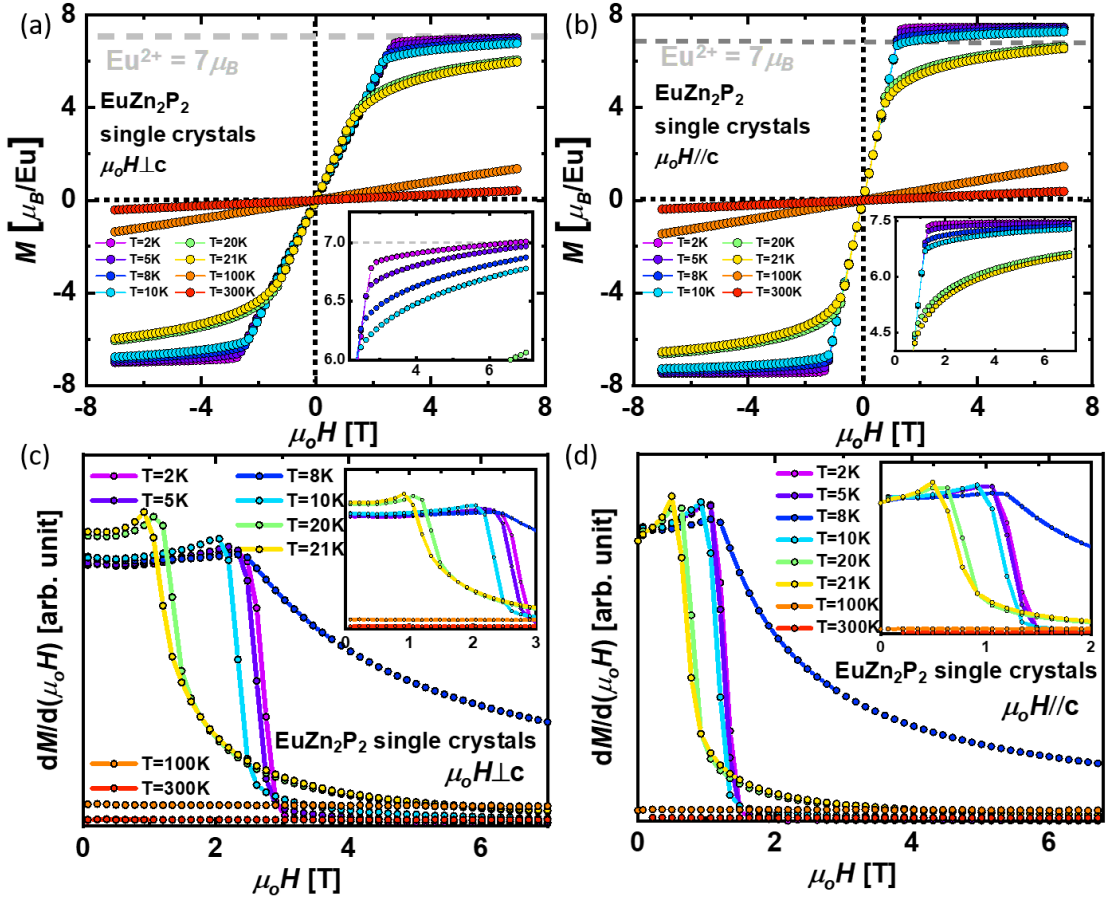

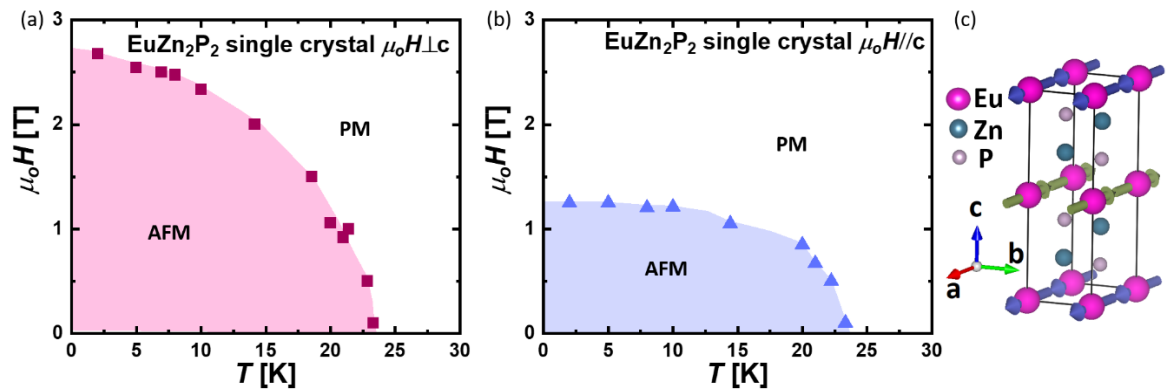

**Figure 5.** Magnetic phase diagram of EuZn$_2$P$_2$ with **(a)** $\mu_o H \perp c$ and **(b)** $\mu_o H \| c$. In both cases, there is an anisotropic AFM state at low fields, followed by a PM state that continuously evolves to a field-polarized state. The transition from AFM to a field-polarized state was determined from the maximum in the M(T) data. **(c)** Proposed A-type antiferromagnetic structure of EuZn$_2$P$_2$.

**Figure 6.** (a) Four probe resistivity data of EuZn$_2$P$_2$ single crystals at $T$=140-400K. (b) Natural logarithm of normalized resistivity as a function of $\frac{1}{T}$ measured on a single crystal of EuZn$_2$P$_2$ along the c-axis. The bandgap extrapolated is $E_g$=0.11eV and is fit to $ln\left(\frac{\rho}{\rho_{T=400K}}\right) = \frac{E_g}{k_B}\frac{1}{T}$.

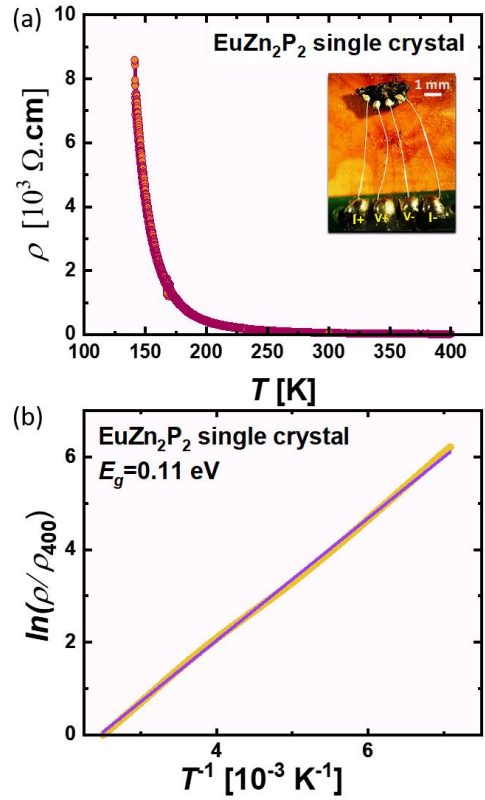

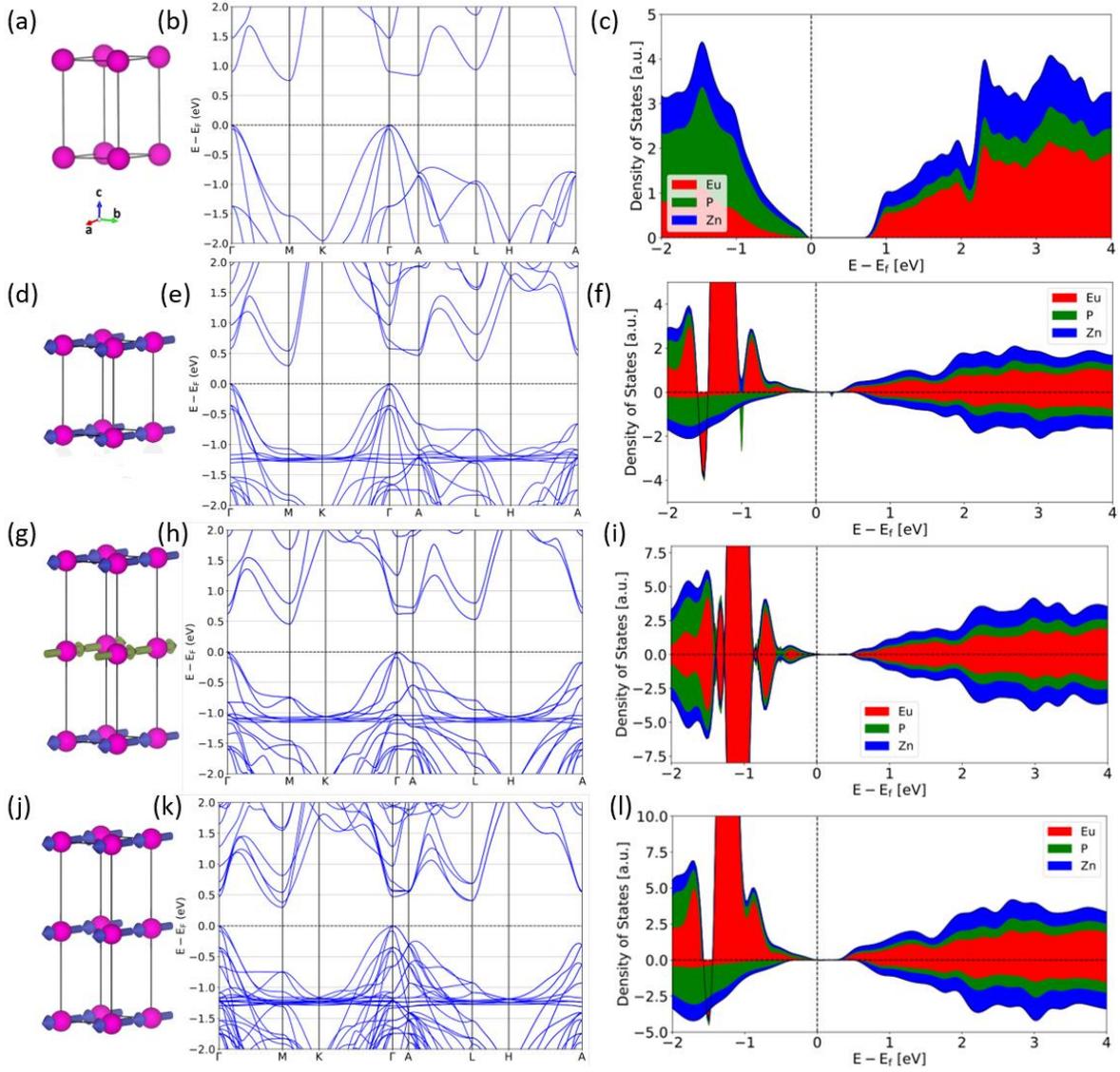

**Figure 7.** Density functional theory (DFT) calculation in the non-magnetic configuration **(a)-(c)**, ferromagnetic configuration (single unit cell) **(d)-(f)**, A-type antiferromagnetic configuration **(g)-(i)**, ferromagnetic configuration (doubled unit cell) **(j)-(l)**. For each case we present the unit cell, band structure, and the density of states.

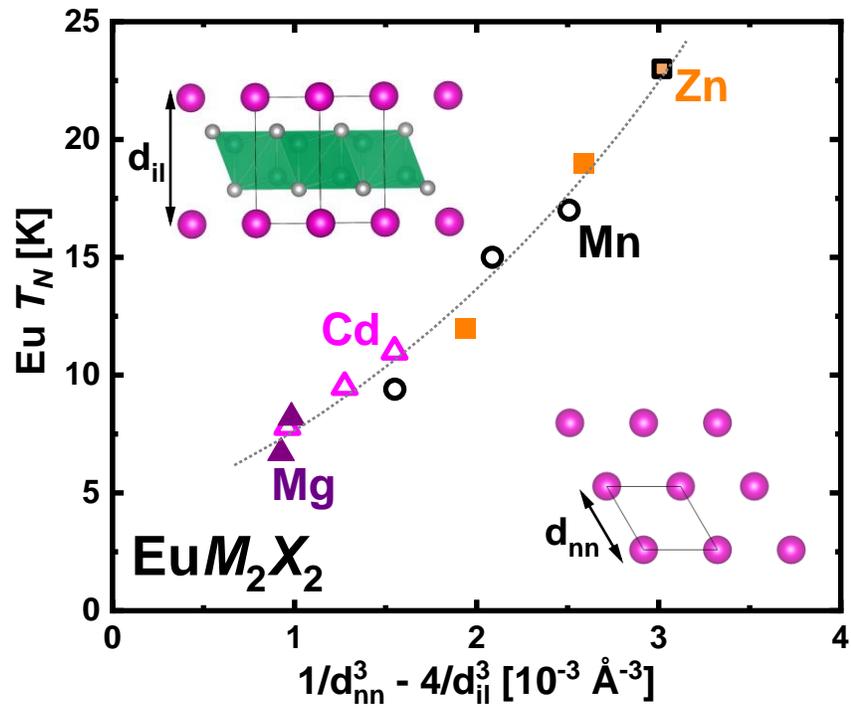

**Figure 8.** The Eu magnetic ordering temperature of Eu$M_2X_2$ materials scales with the inverse cubed distance between intralayer and interlayer Eu-Eu ions within the cationic lattice implying dominate dipolar interactions.

**Table 1.** Crystal data and refinement results for $EuZn_2P_2$

| Formula | $EuZn_2P_2$ |
|---|---|
| Crystal system | Trigonal |
| Space group | $P\bar{3}m1$ (No. 164) |
| FW (g/mol) | 268.96 |
| a (Å) | 4.08497(18) |
| c (Å) | 7.0019(4) |
| V (Å$^3$) | 101.187(11) |
| Z | 1 |
| T (K) | 213 (2) |
| Adsorption coefficient | 27.612 |
| Mo Kα (Å) | 0.71073 |
| Reflections collected / Number of parameters | 180/10 |
| Goodness of fit | 1.188 |
| R[F]$^a$ | 0.0104 |
| $R_w(F_o^2)^b$ | 0.0246 |

$^a$ $R(F) = \Sigma||F_o| - |F_c||/\Sigma|F_o|$
$^b$ $R_w(F_o^2) = [\Sigma w(F_o^2 - F_c^2)^2/\Sigma w(F_o^2)^2]^{1/2}$

**Table 2.** Fractional atomic coordinates and isotropic displacement parameters based on the refined EuZn$_2$P$_2$ structure

| Element | Wyckoff Positions | x | y | z | Occupancy | U$_{iso}$ |
|---|---|---|---|---|---|---|
| Eu | 1a | 1 | 1 | 0 | 1 | 0.00674(10) |
| Zn | 2d | 2/3 | 1/3 | 0.36947(6) | 1 | 0.00791(11) |
| P | 2d | 1/3 | 2/3 | 0.26915(12) | 1 | 0.00637(16) |

**Table 3.** Anistropic displacement parameters based on the refined EuZn$_2$P$_2$ structure

| Element | $U_{11}$ | $U_{22}$ | $U_{33}$ | $U_{23}$ | $U_{13}$ | $U_{12}$ |
|---|---|---|---|---|---|---|
| Eu | 0.00656(11) | 0.00656(11) | 0.00711(13) | 0.000 | 0.000 | 0.00328(5) |
| Zn | 0.00781(13) | 0.00781(13) | 0.0081(2) | 0.000 | 0.000 | 0.00391(6) |
| P | 0.0061(2) | 0.0061(2) | 0.0068(4) | 0.000 | 0.000 | 0.00307(11) |

**Table 4.** Fitting parameters to the $C_p/T$ as a function of $T$ for EuZn$_2$P$_2$ to extract the phonon contribution.

| $s_{D1}$ (oscillator strength/formula unit) | $s_{D2}$ (oscillator strength/formula unit) | $\theta_{D1}$ (K) | $\theta_{D2}$ (K) |
|---|---|---|---|
| 2.35(3) | 2.83(3) | 493(4) | 182(3) |

**Table 5.** Fitting parameters obtained by Curie Weiss analysis from magnetization data of EuZn$_2$P$_2$.

| EuZn$_2$P$_2$ | $\mu_0 H \perp c$ | $\mu_0 H \parallel c$ |
|---|---|---|
| **Range [K]** | 150-300 | 150-300 |
| **C [emu.K.(mol Eu)$^{-1}$.Oe$^{-1}$]** | 9.26 | 7.51 |
| **θ [K]** | 19.2 | 41.9 |
| **χ$_0$ [emu$^{-1}$.(mol Eu).Oe]** | 0.0095 | 0.0002 |
| *p$_{eff}$* | 8.61 | 7.75 |

**Table 6.** Key Eu lattice geometric parameters, observed magnetic ordering temperature, and room temperature resistivity, for a range of known Eu$M_2X_2$ trigonal Zintl materials.

| Eu$M_2X_2$ | Interlayer Eu-Eu distance (Å) | Intralayer Eu-Eu distance (Å) | Eu $T_N$ (K) | $\rho_{ab}$ (mOhm-cm at $T$=300K) | References |
|---|---|---|---|---|---|
| EuZn$_2$P$_2$ | 7.0019 | 4.0845 | 23 | 66 | This Work |
| EuZn$_2$As$_2$ | 7.1810 | 4.2110 | 19 | 75 | 19,30 |
| EuZn$_2$Sb$_2$ | 7.6090 | 4.4938 | 12 | 1.3 | 31,32 |
| EuMn$_2$P$_2$ | 6.9936 | 4.1294 | 17 | 3.50 x 10$^5$ | 35 |
| EuMn$_2$As$_2$ | 7.2250 | 4.2870 | 15 | 200 | 26 |
| EuMn$_2$Sb$_2$ | 7.6740 | 4.5810 | 9.4 | - | 26,27 |
| EuCd$_2$P$_2$ | 7.1790 | 4.3250 | 11 | 24 | 36,37 |
| EuCd$_2$As$_2$ | 7.3500 | 4.4499 | 9.5 | 15 | 38,39 |
| EuCd$_2$Sb$_2$ | 7.7230 | 4.6980 | 7.8 | 12 | 37,40 |
| EuMg$_2$Sb$_2$ | 7.7240 | 4.6950 | 8.2 | - | 41 |
| EuMg$_2$Bi$_2$ | 7.8483 | 4.7724 | 6.7 | 2.3 | 28,29 |